\documentstyle[12pt]{article}
\begin{document}

\title{Same-sign dileptons as a signature for heavy Majorana neutrinos in
hadron-hadron collisions}

\author{F. M. L. Almeida Jr.$^{(1)}$\thanks{marroqui@if.ufrj.br},
Y. A. Coutinho$^{(1)}$\thanks{yara@if.ufrj.br}, 
J. A. Martins Sim\~oes$^{(1)}$\thanks{simoes@.if.ufrj.br},\\
 P. P. Queiroz Filho$^{(2)}$\thanks{pedro@vax.fis.uerj.br} and 
C. M. Porto$^{(3)}$ \\$^{(1)}$Instituto de F\'\i sica\\
Universidade Federal do Rio de Janeiro,\\
$^{(2)}$ Departamento de F\'\i sica Nuclear e Altas Energias\\
Universidade Estadual do Rio de Janeiro,\\Rio de Janeiro\\
$^{(3)}$ Instituto de Ci\^encias Exatas
\\Universidade Federal Rural do Rio
de Janeiro\\Serop\'edica\\ RJ, Brazil\\}

\date{}
\maketitle
\begin{abstract}
\par
We discuss the possibility of same-sign dileptons as a signature for 
Majorana neutrinos. The production mechanism is given by a single
heavy neutrino production and decay
$pp \longrightarrow l^{\pm} N X \longrightarrow l^{\pm} l^{\pm} X'$. Cross
section and distributions are presented for the 
LHC energies.

\end{abstract}
\newpage 
\
In many extensions of the standard model such as left-right models,
SO(10) and $E_6$ models we have the possibility of new heavy neutrinos
\cite{VAL}. Heavy neutrinos are expected to play a central role in the
understanding of the mechanism responsible for small masses for
standard neutrinos. In the near future, direct neutrino detection with
masses of a few TeV are feasible only at CERN high energy proton-proton
collider LHC.
\par
In this paper we turn our attention to the possibility of the production
of a single heavy Majorana neutrino at the LHC energies. The case of pair 
production is also possible and it was studied same time ago\cite{DIC}. 
There is a recent experimental search for pair production of
heavy neutrinos at LEP 1.5 with negative results\cite{ALE}.
We would like to call attention that in most cases it was considered
the possibility of a fourth generation heavy neutrino with 
standard couplings to the $Z^0$. But this is not the only possibility.
If we also have mixing betweeen light and heavy states in the same
family, then the $NNZ$ vertex is suppressed by a factor $\sin^2\theta_{mix}$.
 On the other hand, vertices of the type $\bar \nu NZ$, $eNW$ are suppressed
only by $\sin \theta_{mix}$. We have then the more interesting
possibility of single heavy neutrino production.

\par
In see-saw models the light-to-heavy mixing is too small to give
detectable effects but there are other possibilities. We can have models
with new heavy neutrinos with a mixing angle which is independent of mass ratios. 
This is the case for models with more than one isosinglet
right-handed neutrino\cite{CHN} or in unified theories with B-L
breaking\cite{BUC}.
In order to fix notation we add right-handed neutrino components 
to the standard families,
\begin{equation}
{{\nu^0_e \choose e}_L, \nu^0_{eR}, e_R;
\qquad {\nu^0_{\mu} \choose \mu}_L, \nu^0_{\mu R}, \mu_R; ...}
\end{equation}

\
\par
For the eigenstates,
\begin{eqnarray}
\chi_{\nu i} & = &\nu^0_{iL}+(\nu^0_{iL})^c \nonumber\\
\omega_{\nu i} & = & \nu^0_{iR}+(\nu^0_{iR})^c
\end{eqnarray}
 the mass matrix has the form:
\begin{equation}
\left(
\begin{array}{cc}
0 & D \\
D^T & M_R
\end{array}
\right)
\end{equation}
\par
One can diagonalize this matrix by a unitary matrix $U$, such that 
$\tilde M=U^TMU=$ (diagonal, real).
The mass eigenstates are written as combinations of interacting states:
\begin{eqnarray}
\chi_{\nu i} & = & \sum_{k=1}^{2n} U_{ik} \rho_k ,\qquad i=1,2,...,n \nonumber\\
\omega_{\nu i} & = & \sum_{k=1}^{2n} U_{ik} \rho_k,\qquad i=n+1,...,2n
\end{eqnarray}
where ``n'' is the family index.
The mass matrix can be diagonalized by blocks\cite{SCH} through the matrix
product:
\begin{equation}
U=U'V=\left[
\begin{array}{cc}
(1-{1\over 2}\xi^*\xi^T)V_1 & \xi^*V_2 \\
-\xi^TV_1 & (1-{1\over 2}\xi\xi^*)V_2
\end{array}
\right] + O(\xi^3)
\end{equation}
where $\xi=DM_R^{-1}$ and $V_1$, $V_2$ are unitary matrices. In the see-saw
mechanism $M_R$ is very large  and the standard neutrinos are very light.
In this case the mixing parameters become very small. 
An alternative
point of view is to have some singular mass matrix\cite{CHN},\cite{TOM}. This implies
in a nearly zero eigenvalue for the known neutrinos masses and the mixing 
angles are free parameters, fixed from phenomenological constraints. Then
the mixing parameters for light to heavy leptonic transitions can be
large even if the neutrino mass spectrum contains very light and
heavy states.
\par
The lagrangians relevants for our purpose are:
\begin{eqnarray}
{\cal L}_{cc} & = & -{g\over 2\sqrt2}W_{\mu}{\sum_{i=1}^{n}}
\bar l_i \gamma^{\mu}(1-\gamma^5)\bigl[{ \sum_{j=1}^{n}}
\bigl[(1 - {1\over 2}\xi^* \xi^T)V_1\bigr]_{ij}\nu_j 
\nonumber\\
& + &  {\sum_{j=n+1}^{2n}} (\xi^*)_{ij} N_j\bigr] 
\end{eqnarray}
and
\begin{eqnarray}
{\cal L}_{nc} & = & -{g\over 4\cos\theta_W}Z_{\mu}{\sum_{i=1}^{n}}
{\sum_{j,k=n+1}^{2n}}\bar N_k \gamma^{\mu}(1-\gamma^5)
(\xi^*V_2)^{\dagger}_{ki}(\xi^*V_2)_{ij}N_j
\end{eqnarray}
where we call $l_i$ the usual charged leptons, $\nu_i$ the light
neutrinos and $N_i$ the heavy neutrinos.
\par
In order to have a general situation we consider the mixing of an
``electron-type light neutrino'' and a new heavy Majorana neutrino
with a mixing parameter fixed only by phenomenological 
consequences\cite{SIM}. The most stringent bounds come from the
LEP data on the $Z^0$  properties\cite{NAR}. They are consistent with
a bound of the order of $sin^2\theta_{mix} < 10^{-2}-10^{-3}$.
\par

Single heavy neutrino can be produced via:
\par
\begin{equation}
p+p \longrightarrow e^- \bar NX \longrightarrow e^- e^- W^+ X
\end{equation}
\par
We can also have same-sign electrons in pair production of heavy Majorana neutrinos
\begin{equation}
p+p \longrightarrow  N \bar N\longrightarrow e^- e^- W^+ W^+ Z 
\end{equation}
but supressed by a term $\sin^4\theta_{mix}$. If the heavy neutrino comes from the muon
family we will have a same-sign dimuon in the final state.
\par
For reaction (8) one straightforward calculate\cite{DJO}
Drell-Yan cross sections. For Majorana neutrinos we have the charged 
current decay 
$N \longrightarrow l^{\pm} W^{\mp}$. This gives a very clear final signature with two same sign charged leptons in the final state. The final signature for heavy neutrino decay with the
highest branching ratio will be given by the hadronic channel $N \longrightarrow l^{\pm}
W^{\mp} \longrightarrow l^{\pm} +$ hadrons. This is an interesting 
signature because there is no missing energy in the final state, and
hadronic jet must have an invariant mass of $M^2_W$.
\par
We have calculated cross sections and distributions using standard
Monte Carlo methods. We show in figure 1 the total cross section 
at LHC energies for
single heavy neutrino production. We employ the Gluck-Reya-Vogt parton distributions functions\cite{GLU}. If we consider an annual luminosity
of 100 $fb^{-1}$ and a rate of 10 events/year as feasible we see that
LHC can detect Majorana neutrinos with mass up to 1.4 TeV. Of course
this is tied to a mixing angle of $10^{-2}$. For the more stringent bound on the 
mixing angle $\sin^2\theta_{mix}=10^{-3}$ we can reach masses up to 800 GeV.
If no such neutrino is found this result can improve considerably the forbidden mass-angle region. There is an important effect on the total cross section due to the phase space for single heavy neutrino production. This is most clearly show in the variable 
$\tau=x_{1}x_{2}={\displaystyle{ \hat s \over s}}$. The dominant region is at small $\tau$ and the kinematical 
threshold for single production at $M^2_{N}$ implies a larger cross section relative
to pair production with threshold at $4M^2_{N}$. This is shown in figure 2.
 In figure 3 we show the primary lepton angular distribution. We call $\theta_1$ the
angle of the outgoing lepton with the beam. For lighter neutrino mass it is peaked 
in the beam direction. As the heavy neutrino mass increases, this effect becomes
smaller. The secondary lepton angular distribution follows the same pattern. In order to check if the cross section is still significative after
an angular cut in the beam direction, we impose
an angular cut for all the final particles (leptons and hadrons).
The fraction of the signal outside this cone is shown in figure 4.
\par
The most favored final signature will be $l^{\pm} l^{\pm}$ hadrons, with the
hadronic jet showing a high $P_T$ distribution peak around 
$\displaystyle{ M_N\over 2}$ as shown in figure 5.
\par
In conclusion we have shown that the same-sign dileptons at the LHC energies can give a clear signal for heavy Majorana neutrinos with masses
in the region of a few hundreds of GeV up to 1.4 TeV.
 The hadronic $P_T$ distribution gives a precise determination of the
heavy neutrino mass.

\vskip 2cm
\vfill\eject
Acknowledments 
\par
This work was partially supported by CNPq, FINEP and FUJB.
\vskip 1cm
\vspace{1cm}
\LARGE
Figure Captions
\normalsize
\begin{itemize}
\item 1. Total cross sections for pair and single heavy Majorana
neutrinos at $\sqrt s =14$ TeV and $\sin^2 \theta_{mix}=10^{-2}$.
\item 2.a Tau distribution for a single heavy Majorana neutrino.
\item 2.b Tau distribution for a pair of heavy Majorana neutrinos.
\item 3. Angular distribution of the primary lepton relative to the
beam axis.
\item 4. Total cross sections with an angular cut on the dileptons
and hadronic jet. 
\item 5. Hadronic $P_T$ distributions for several heavy neutrino masses.
\end{itemize}

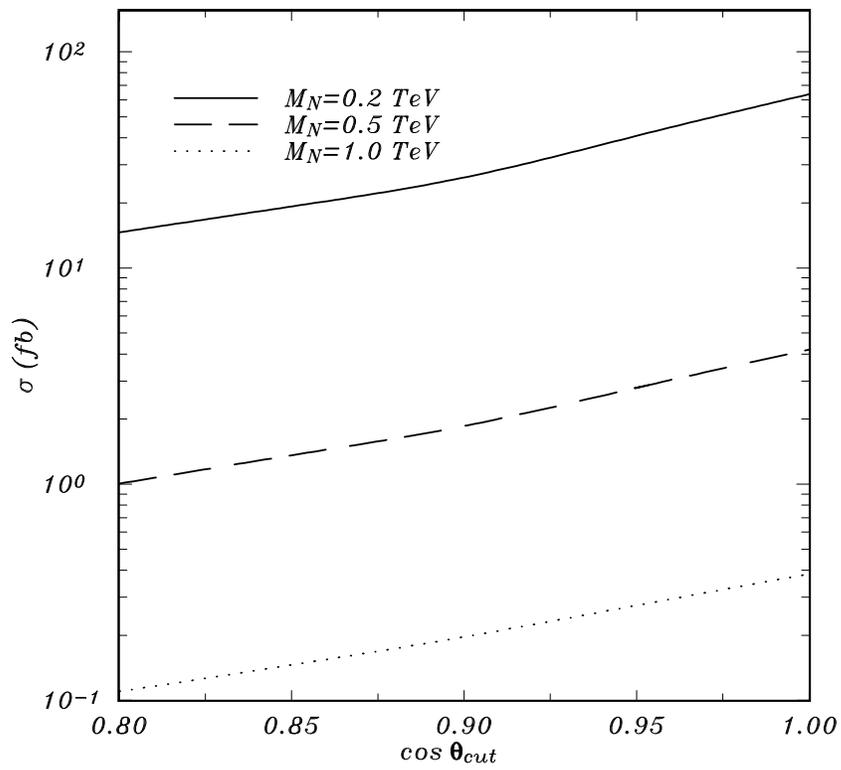
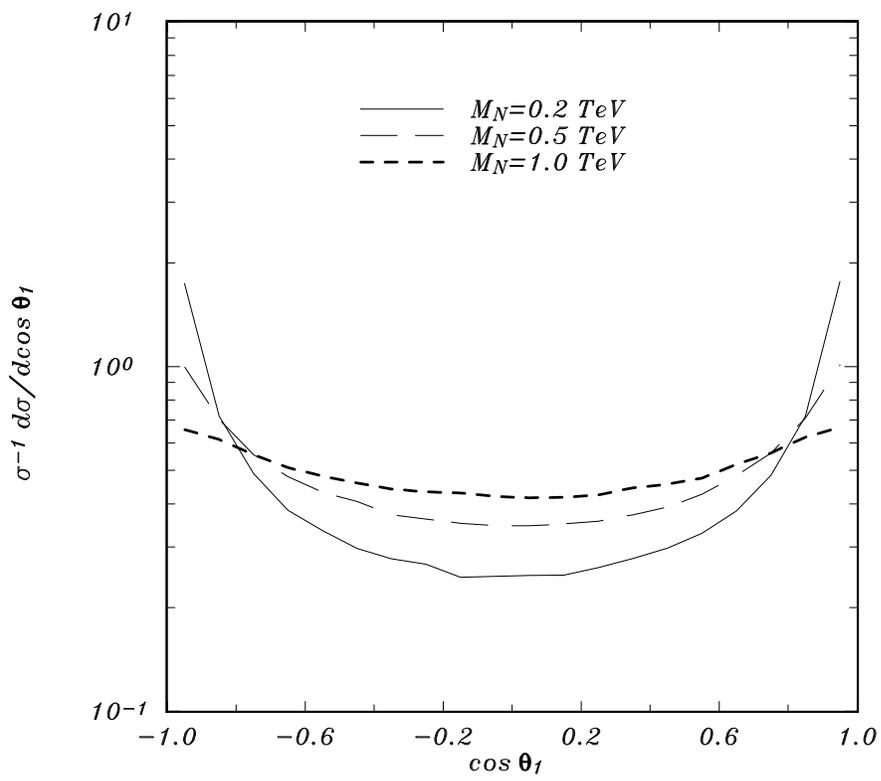
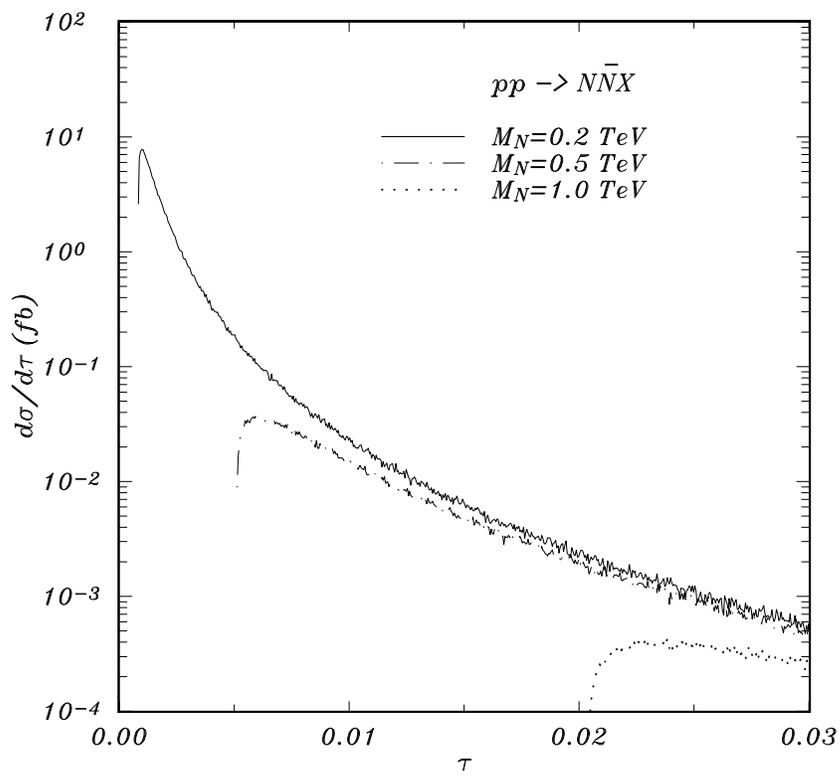
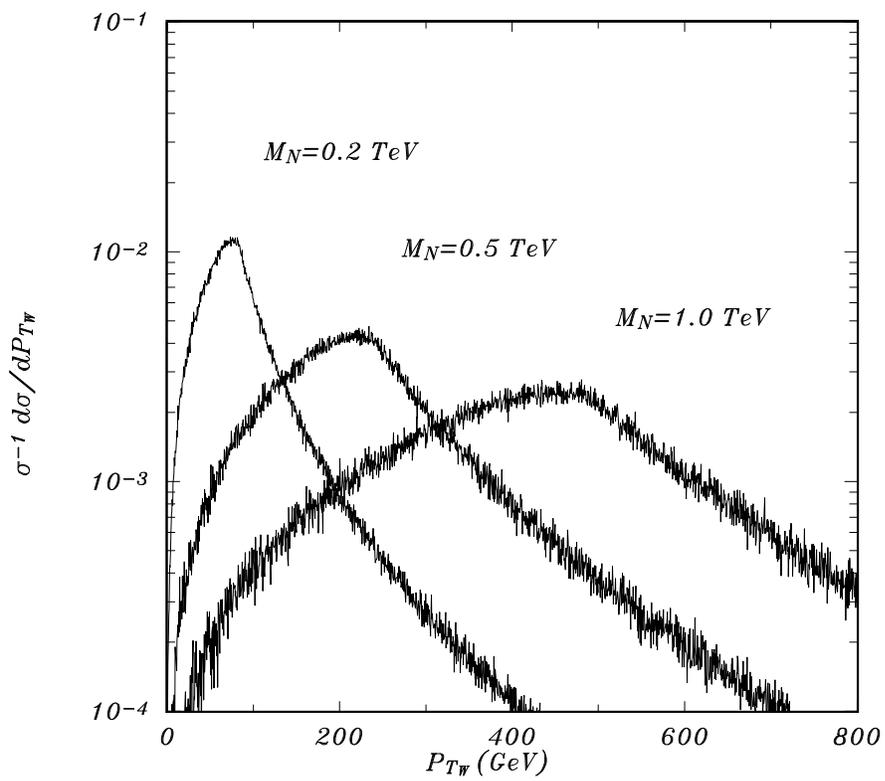
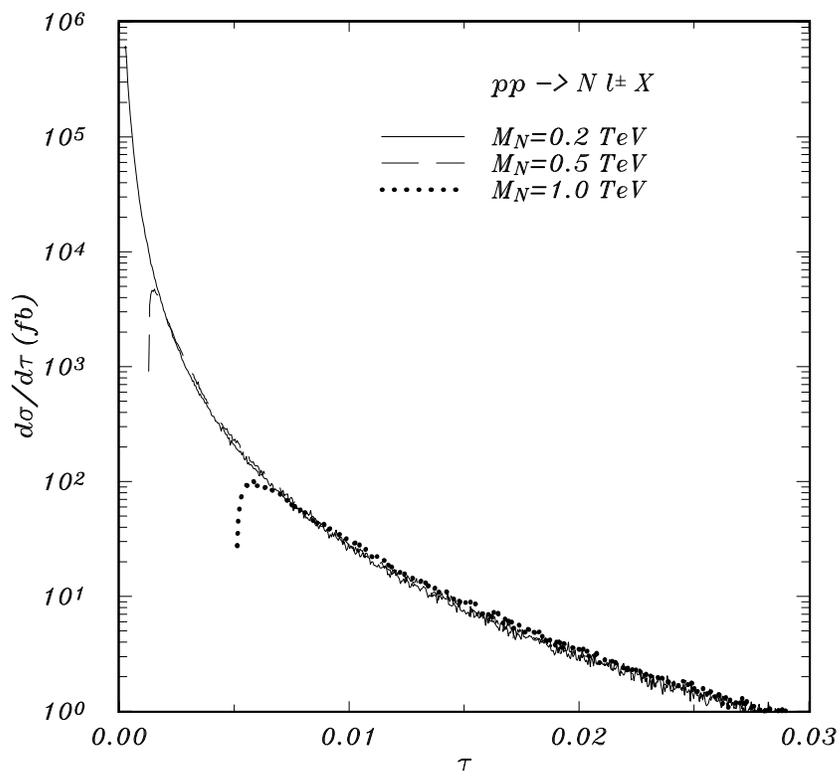
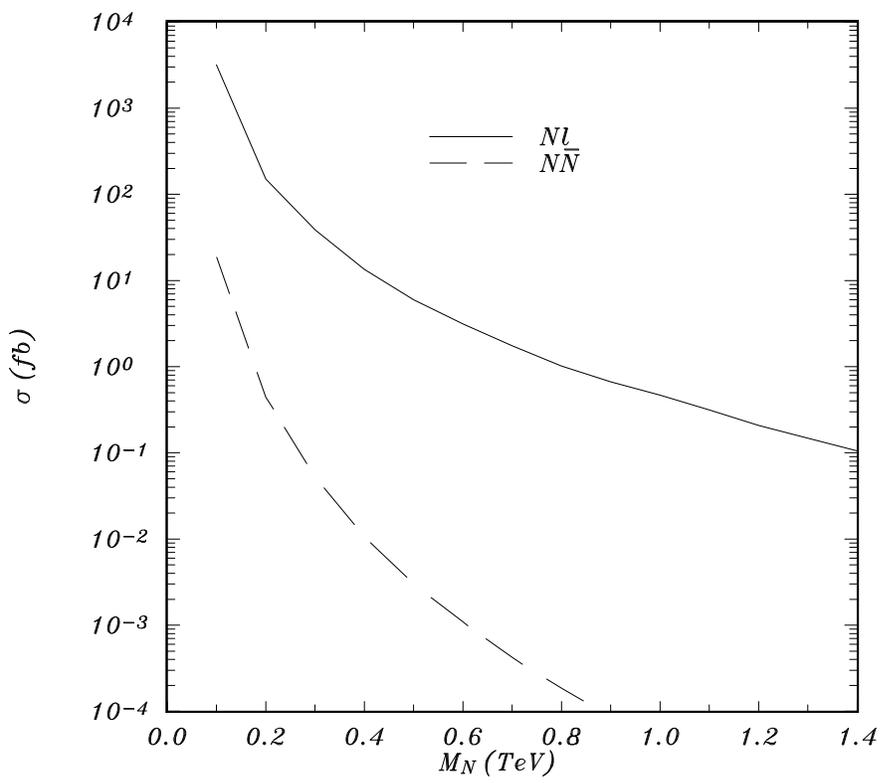
\end{document}